\documentclass{aa}

\usepackage{txfonts}
\usepackage{placeins}           
\usepackage{orcidlink}          
\usepackage{hyperref}
\usepackage[utf8]{inputenc}
\usepackage{amssymb,amsmath,verbatim,mathtools,needspace,enumitem,etoolbox,graphicx,physics,microtype,ulem,breqn}
\usepackage{xspace} 
\usepackage{booktabs}
\usepackage{upgreek} 
\usepackage{acronym}
\usepackage{url}

\newcommand\cd{d$^{-1}$\,}
\newcommand\kms{km\,s$^{-1}$\,}
\newcommand{\logg}{$\log g$}
\newcommand{\vsini}{$v \sin i$}
\newcommand{\teff}{$T_{\rm eff}$}
\newcommand{\Msun}{$\rm M_{\odot}$}

\usepackage{color}
\definecolor{todo}{rgb}{0.89,0.0,0.13}

\usepackage{color}
\definecolor{review}{RGB}{191, 66, 245}

\usepackage{color}
\definecolor{green(new)}{RGB}{50, 200, 60}

\usepackage{color}
\definecolor{cadetgrey}{rgb}{0.57, 0.64, 0.69}

\usepackage{color}
\definecolor{oror}{RGB}{0,150,0}

\usepackage{amsmath}

\begin{document}

\title{EL CMi: confirmation of triaxial pulsation theory}

\author{G. Handler \orcidlink{0000-0001-7756-1568} \inst{\ref{CAMK}}
\and S. A. Rappaport \orcidlink{0000-0003-3182-5569} \inst{\ref{MKI}}
\and D. Jones \orcidlink{0000-0003-3947-5946} \inst{\ref{Can1},\ref{Can2}}
\and A. Miszuda \orcidlink{0000-0002-9382-2542} \inst{\ref{CAMK}}
\and{M. Omohundro}\inst{\ref{Zoo}}
\and{R. Jayaraman} \orcidlink{0000-0002-7778-3117} \inst{\ref{MKI}}
\and{R. Gagliano} \orcidlink{0000-0002-5665-1879} \inst{\ref{Glen}}
\and \\ {J. Fuller} \orcidlink{0000-0002-4544-0750} \inst{\ref{CIT}}
\and{D. W. Kurtz} \orcidlink{0000-0002-1015-3268} \inst{\ref{Mahi},\ref{UCL}}
\and{J. Munday} \orcidlink{0000-0002-1872-5398} \inst{\ref{UoW}}
\and{H.-L. Chen} \inst{\ref{YO},\ref{YKL}}
\and{B. P. Powell \orcidlink{0000-0003-0501-2636}} \inst{\ref{NGS}}
\and{V. B. Kostov \orcidlink{0000-0001-9786-1031}} \inst{\ref{NGS},\ref{SETI}}}

\institute{Nicolaus Copernicus Astronomical Center, Polish Academy of Sciences, ul. Bartycka 18, PL-00-716 Warszawa, Poland \\
\email{gerald@camk.edu.pl} \label{CAMK} 
\and Department of Physics and Kavli Institute for Astrophysics and Space Research, Massachusetts Institute of Technology, 77 Massachusetts Ave, Cambridge, MA 02139, USA \label{MKI}
\and Instituto de Astrof\'isica de Canarias, 38205 La Laguna, Spain \label{Can1}
\and Departamento de Astrof\'isica, Universidad de la Laguna, 38206 La Laguna, Tenerife, Spain \label{Can2}
\and Citizen Scientist, c/o Zooniverse, Dept., of Physics, University of Oxford, Denys Wilkinson Building, Keble Road, Oxford OX1 3RH, UK \label{Zoo}
\and Amateur Astronomer, Glendale, AZ 85308, USA \label{Glen}
\and TAPIR, Mailcode 350-17, California Institute of Technology, Pasadena, CA 91125, USA \label{CIT}
\and Centre for Space Research, North-West University, Mahikeng 2745, South Africa \label{Mahi}
\and Jeremiah Horrocks Institute, University of Central Lancashire, Preston PR1 2HE, UK \label{UCL}
\and Department of Physics, Gibbet Hill Road, University of Warwick, Coventry CV4 7AL, UK \label{UoW}
\and Yunnan Observatories, Chinese Academy of Sciences (CAS), Kunming 650216, People's Republic of China \label{YO}
\and International Centre of Supernovae, Yunnan Key Laboratory, Kunming 650216, People's Republic of China\label{YKL}
\and NASA Goddard Space Flight Center, 8800 Greenbelt Road, Greenbelt, MD 20771, USA\label{NGS}
\and SETI Institute, 189 Bernardo Ave, Suite 200, Mountain View, CA 94043, USA\label{SETI}
}
\abstract{
Triaxial pulsators are a recently discovered group of oscillating stars in close binary systems that show pulsations around three axes at the same time. It has recently been theoretically shown that new types of pulsation modes, the Tidally Tilted Standing (TTS) modes, can arise in such stars. Here, we report the first detection of a quadrupole TTS oscillation mode in the pulsating component of the binary system EL CMi following an analysis of {\it TESS} space photometry. Two dipole oscillations around different axes in the orbital plane are present as well. In addition, the binary system is characterized using new radial velocity measurements, \textsc{phoebe} as well as simultaneous spectral energy distribution and light curve modeling. The pulsating primary component has properties typical of a $\delta$ Scuti star but has accreted and is still accreting mass from its Roche Lobe filling companion. The donor star is predicted to evolve into a low-mass helium white dwarf. EL CMi demonstrates the potential of asteroseismic inferences of the structure of stars in close binaries before and after mass transfer and in three spatial dimensions.
}


\maketitle

\section{Introduction}

The development of high-precision photometric space missions such as {\it CoRoT} \citep{2006cosp...36.3749B}, {\it Kepler} \citep{2010ApJ...713L..79K}, {\it TESS} \citep{ricker}, {\it CHEOPS} \citep{2021ExA....51..109B} and {\it PLATO} \citep{2014ExA....38..249R} has been primarily motivated by the quest for extrasolar planets by detecting transits and for the characterization of those planets. However, besides the orbital inclination and information on limb darkening the transit method just yields the ratio of the planetary to stellar radius. To determine absolute dimensions, knowledge about the size of the planetary host star is required.

This is where asteroseismology comes into play. Asteroseismology determines the interior structure of stars by using stellar oscillations as seismic waves \citep[e.g.,][and references therein]{2021RvMP...93a5001A,2022ARA&A..60...31K} and is at the same time capable of constraining the absolute parameters of its target stars \citep[e.g.,][]{2013ApJ...767..127H}. As asteroseismic investigations themselves also greatly benefit from high accuracy photometric data, almost all of the planet search space missions have an asteroseismology program as well \citep[e.g.,][]{2010PASP..122..131G}.

The outcome from those missions in terms of asteroseismic research has been rich. To mention just a few results simply from \texttt{TESS}, more than 150\,000 oscillating red giant \citep{2021ApJ...919..131H} and over 100\,000 pulsating A-F type stars \citep{2024ApJ...972..137G} have been discovered, period spacing patterns permitting asteroseismology of dozens of $\delta$ Scuti stars for the first time \citep{2020Natur.581..147B} were detected, low-frequency oscillations in massive supergiant stars appear to be ubiquitous \citep{2019NatAs...3..760B}, and the interiors of individual massive stars could be studied \citep{2023NatAs...7..913B}.

Obviously, the availability of data of unprecedented quality also facilitated the discovery of unexpected phenomena, such as a star that stopped pulsating \citep{2025MNRAS.536.2103K} or stars in close binary systems that have their pulsation axis tilted with respect to their rotation axis due to the gravitational pull of their companion star \citep{2020NatAs...4..684H}. Soon afterwards, several more of those systems were reported \citep{2020MNRAS.494.5118K,2021MNRAS.503..254R,2022MNRAS.510.1413K,2024MNRAS.533.2705J}. Whereas the pulsating star in the latter binaries was of the $\delta$ Scuti type, a tidally tilted subdwarf B pulsator was also discovered \citep{2022ApJ...928L..14J}. From the point of view of theory, the pulsations of these objects were explained via tidal mode coupling that both tilts the pulsation axes and confines the pulsations to certain regions of the star \citep{2020MNRAS.498.5730F}.

More recently, two of the tidally tilted pulsators have been found to be ``triaxial'' pulsators with dipole pulsations about three orthogonal axes, including the tidal axis \citep{2024MNRAS.528.3378Z,2024ApJ...975..121J}. Based on those discoveries, \cite{2025ApJ...979...80F} developed a theoretical description of how these modes of pulsation may appear in triaxially deformed stars. Henceforth, these will be called ``Tidally Tilted Standing" modes (TTS; sometimes ``Fuller modes" for short). These are modes that arise when tidal, centrifugal, and Coriolis perturbations couple modes of equal spherical degree \citep{2025ApJ...979...80F}. Such modes turn out to be triaxial for dipole configurations, while for higher multipole modes, their geometry is more complex. Since these modes typically have nearly equal contributions from spherical harmonics of spherical degrees $+m$ and $-m$, this means the modes are primarily standing modes that do not propagate around the rotation axis. Instead they have a pattern that is nearly stationary with respect to the companion star.

Furthermore, based on this theory \cite{2025ApJ...979...80F} predicted how quadrupole triaxial pulsation modes would manifest themselves observationally and provided the theoretical expressions for quadrupole and octupole TTS pulsations. Here, we report the discovery of a textbook example of a pulsator with two dipole modes around different axes in the orbital plane plus a quadrupole ``Fuller mode", occurring on the primary component of the eclipsing binary system EL CMi.  

\section{Observations and Analysis}

\subsection{Photometry} \label{sec:phot}

EL CMi (TIC 321254859) has been observed with \texttt{TESS} in Full Frame Images taken during Sectors 7 (January 2019), 34 (January 2021), 61 (February 2023), and 88 (January 2025) with cadences of 1800, 600, 200 and 200 s, respectively. \texttt{TESS-SPOC} light curves are available for the first two sectors, whereas our own \texttt{lightkurve} reductions were made for the Sector 61 and 88 data. All light curves show some drifts in the mean brightness which were filtered out and segments of poor data quality were removed.  EL CMi was initially identified as a high-scoring output of a neural network designed to find eclipses in TESS light curves, described further in \citet{2021AJ....161..162P} and \citet{2025arXiv250605631K}. Upon manual review, the presence of tidally tilted or triaxial pulsations was suspected in the Sector 61 data. Consequently, the system was subjected to closer examination regarding the periodic content in its light curves.

To perform frequency analyses, the {\texttt Period04} software package \citep{2005CoAst.146...53L} was used. This package applies Fourier Analysis and simultaneous multi-frequency sine-wave fitting. Advanced options such as the calculation of optimal light-curve fits for multiperiodic signals including harmonic, combination, and equally spaced frequencies are available. Fits composed of multiple sine waves are subtracted from the data and the residuals examined for the presence of further periodicities.

\texttt{TESS} data come in different flavors, in particular Simple Aperture Photometry (SAP, \citealt{2010SPIE.7740E..23T,2020ksci.rept....6M}) and Pre-search Data Conditioning SAP (PDCSAP, \citealt{2012PASP..124..985S,2012PASP..124.1000S}) fluxes. The latter are detrended using Co-trending Basis Vectors and are often cleaner than SAP fluxes. On the other hand, the detrending procedure can distort light curves with intrinsic sharp features, such as eclipses, and therefore the choice of the optimal data version sometimes depends on the goals and demands of the scientific analysis to be carried out.

The eclipses in the Sector 7 and 34 SAP data for EL CMi appear shallower than in the PDCSAP and in the Sector 61 and 88 data. This can be explained by blending of nearby stars into the photometric apertures as the corresponding {\it CROWDSAP} parameters (0.8784 and 0.8959, respectively) imply. Therefore, the analysis of the binary light curve is based on the Sector 34 PDCSAP data that should no longer be contaminated by the companion. On the other hand, the Fourier noise level in the range of the pulsation frequencies is considerably lower (by about 30\%) in the SAP data compared to the PDCSAP light curves. Consequently, the SAP data were used for pulsational frequency detection, with the caveat that their amplitudes may be systematically lower than in reality. However, this is of no relevance to the results of the analyses performed on those data.

\begin{figure}
\includegraphics[width=\linewidth,viewport=40 40 395 273]{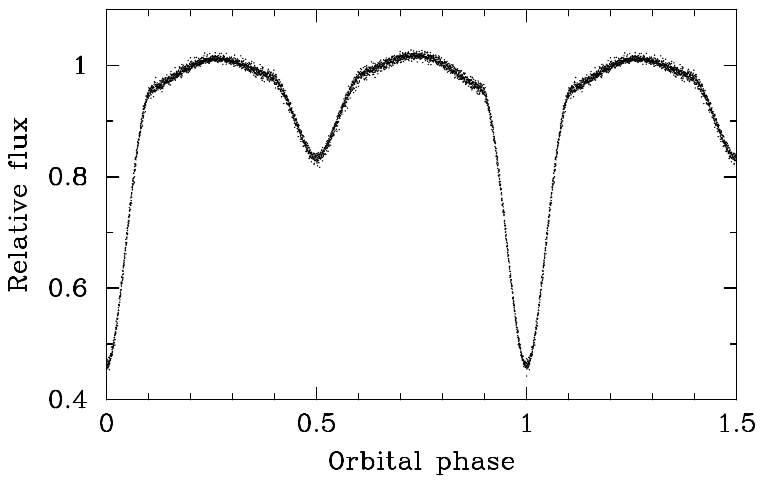}
\caption{The phase folded, pulsation removed \texttt{TESS} Sector 34 PDCSAP light curve of EL CMi.}
\label{elcmiecl}
\end{figure}

In a first step, the ephemeris for the times of primary minimum from the \texttt{TESS} data was determined by computing a multiharmonic fit to the light curves merged into 1800-s bins to give them all equal weight. This ephemeris resulted as:
\begin{displaymath}
	T_I=2459229.2182(2)+1.05384979(7)E,
\end{displaymath}
where the zeropoint in time relates to the center of the first primary eclipse in the Sector 34 data. The corresponding orbital phase curve is shown in Fig.\,\ref{elcmiecl}.

Turning to the analysis of the pulsations, only the Sector 34, 61 and 88 data were used as the pulsation frequencies are above the Nyquist limit of the Sector 7 data. An essential question in the quest for the detection of periodic signals in a data set is where to stop the analysis. To this end, several criteria have been proposed. Here, the methodology originally put forward by \citet{1993A&A...271..482B} was adopted. In this framework, a Fourier peak must exceed the mean amplitude in the amplitude spectrum by a factor of 4 in the local frequency domain to be considered significant. For space-based data, this can however lead to an overinterpretation of the periodic content \citep{2014MNRAS.439.3453B}. For \texttt{TESS} data, \cite{2021AcA....71..113B} have proposed stricter detection criteria depending on the sampling of the data set. Interpolating their values, signal to noise thresholds of 4.8 for the Sector 34 data as well as 5.0 for the Sector 61 and 88 data, respectively, were adopted that correspond to False Alarm Probabilities of 0.1\,\%. The mean amplitude of the noise was evaluated in a 5\,\cd wide window centered at the frequency under consideration.

\begin{figure}
\includegraphics[width=\linewidth,viewport=40 30 385 465]{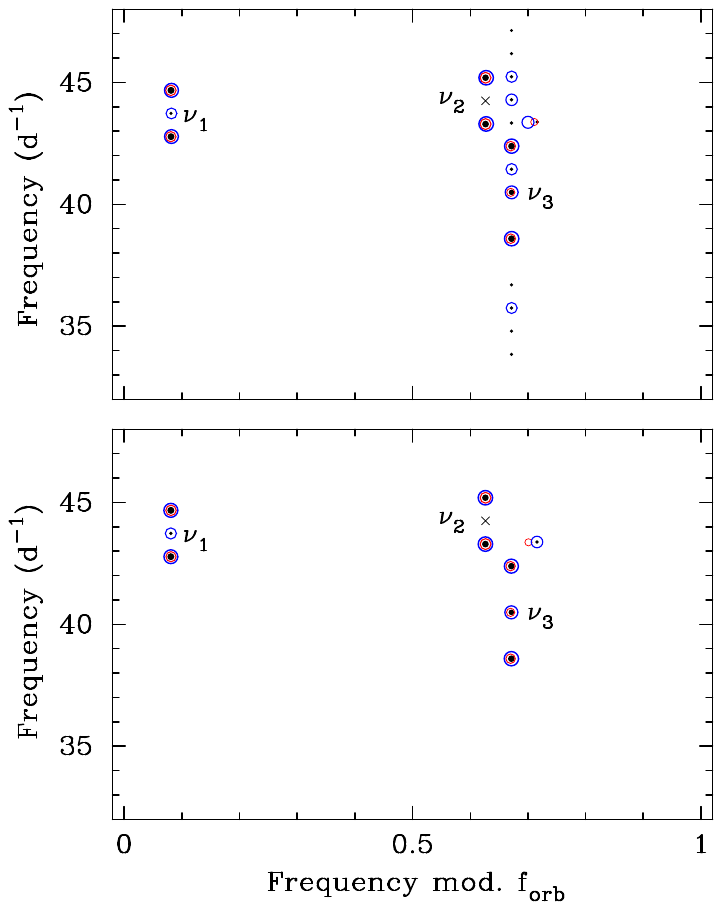}
\caption{\'Echelle Diagram for EL CMi from Sectors 34 (black), 61 (red) and 88 (blue), respectively. Top: all data. Bottom: primary eclipses removed from light curve. The sizes of the plot symbols relate to the logarithm of pulsation amplitude; the cross marks the expected location of the centroid of mode $\nu_2$.}
\label{elcmiech}
\end{figure}

\begin{figure}
\includegraphics[width=\linewidth,viewport=35 70 385 460]{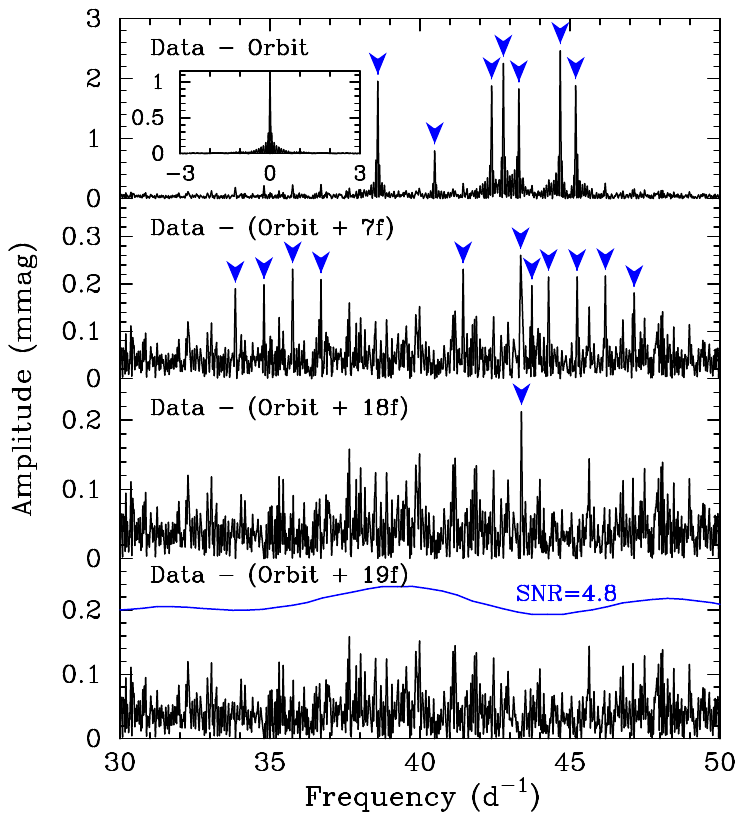}
\caption{Fourier Transforms of the Sector 34 \texttt{TESS} data for EL CMi with subsequent prewhitening steps (first, only the orbital harmonics, then the orbital harmonics plus seven, 18 and 19 pulsation frequencies indicated by the blue arrows, respectively) indicated. The inset in the top panel is the spectral window function of this data set. The level of the SNR curve is applicable for all of these amplitude spectra but is plotted only once for clarity.}
\label{elcmi_fts}
\end{figure}

The pulsation frequency detection process was performed with eclipses accounted for by a multiharmonic fit. Then the pulsation frequencies were added to a fit that includes both the eclipses and the pulsations simultaneously. Once no significant signals were left, this process was stopped. At this point, an \'Echelle Diagram was constructed with respect to the orbital frequency \citep[see][for more details]{2022ApJ...928L..14J}, to evaluate which of the detected signals could form multiplets. The frequencies of confirmed multiplet members were then fixed to their predicted values using the orbital frequency, and the fit to the complete light curve was optimized and re-evaluated to check whether no additional signals had been left behind. The resulting lists of frequencies are given in Table\,\ref{elcmi_frqs}, and the resulting \'Echelle Diagram is shown in Fig. \ref{elcmiech}, upper panel.
The pre-whitening process is illustrated in Fig. \ref{elcmi_fts}. 

\begin{table*}
\centering
\caption{Pulsation related frequencies of TIC 321254859 = EL CMi.}
\scriptsize
\begin{tabular}{l|cccr|cccr|cccr} 
\hline
\hline
ID & Frequency & Ampl. & Phase & S/N & Frequency & Ampl. & Phase & S/N & Frequency & Ampl. & Phase & S/N\\ 
& (\cd) & (mmag) & (rad) & ~ & (\cd) & (mmag) & (rad) & ~  & (\cd) & (mmag) & (rad) & \\
\hline
 & \multicolumn{4}{c}{Sector 34} &  \multicolumn{4}{c}{Sector 61} &  \multicolumn{4}{c}{Sector 88}\\
\hline
\hline
$\nu_1-f_{orb}$  & 42.7778(07) & 2.25 & 2.48(03) & 56.4 & 42.7777(07) & 1.92 &  2.64(03)  &  30.1 & 42.7785(11) & 1.09 &  4.38(05) &  23.9 \\
$\nu_1$          & 43.7267(78) & 0.19 & 2.93(36) &  4.8  & {\it 43.7266} & -- & -- & -- & 43.7274(74) & 0.16 &  4.18(31) &   3.8\\
$\nu_1+f_{orb}$  & 44.6756(06) & 2.44 & 2.46(03) & 61.7 & 44.6755(05) & 2.34 &  2.65(02)  &  33.8 & 44.6763(10) & 1.20 &  4.35(04)&  27.3\\ [0.5ex]
$\nu_2-f_{orb}$  & 43.2952(08) & 1.95 & 5.17(04) & 46.6 & 43.2952(04) & 3.10 &  3.60(02)  &  46.3 & 43.2961(07) & 1.73 &  5.70(03) &  39.7\\
$\nu_2$          & {\it 44.2441} & -- & -- & -- & {\it 44.2441} & -- & -- & -- & {\it 44.2450} & -- & -- & --\\
$\nu_2+f_{orb}$  & 45.1930(08) & 1.81 & 2.00(04) & 45.3 & 45.1930(04) & 2.86 &  0.46(02)  &  41.2 & 45.1939(07) & 1.69 &  2.47(03) &  37.5\\ [0.5ex]
$\nu_3-7f_{orb}$ & 33.8489(81) & 0.18 & 5.35(37) &  4.5  & -- & -- & -- & --  & -- & -- & -- & --\\
$\nu_3-6f_{orb}$ & 34.7978(77) & 0.19 & 5.51(35) &  4.6  & -- & -- & -- & --  & -- & -- & -- & --\\
$\nu_3-5f_{orb}$ & 35.7467(74) & 0.20 & 5.31(34) &  4.7  & -- & -- & -- & -- & 35.7470(65) & 0.18 &  2.24(28)
 &   3.5 \\
$\nu_3-4f_{orb}$ & 36.6956(75) & 0.20 & 5.28(34) &  4.5  & -- & -- & -- & --  & -- & -- & -- & --\\
$\nu_3-2f_{orb}$ & 38.5934(08) & 1.96 & 5.16(03) & 40.3 & 38.5922(10) & 1.32 &  2.31(04)  &  22.4 & 38.5937(08) & 1.56 &  2.01(03) &  28.2\\
$\nu_3$          & 40.4912(19) & 0.78 & 5.25(09) & 16.1 & 40.4900(25) & 0.51 &  2.32(11)  &   8.7 & 40.4915(20) & 0.60 &  2.06(08) &  11.4\\
$\nu_3+f_{orb}$  & 41.4401(66) & 0.23 & 2.52(30) &  5.0  & -- & -- & -- & -- & 41.4404(51) & 0.23 &  5.42(22) &   4.6\\
$\nu_3+2f_{orb}$ & 42.3890(08) & 1.90 & 2.04(04) & 44.8 & 42.3878(09) & 1.32 &  5.43(04)  &  21.0 & 42.3893(08) & 1.44 &  5.11(04) &  30.0\\
$\nu_3+3f_{orb}$ & 43.3379(63) & 0.24 & 2.09(29) &  5.7  & -- & -- & -- & --  & -- & -- & -- & --\\
$\nu_3+4f_{orb}$ & 44.2868(69) & 0.22 & 2.27(31) &  5.5  & -- & -- & -- & -- & 44.2871(47) & 0.25 &  5.10(20) &   5.9\\
$\nu_3+5f_{orb}$ & 45.2357(68) & 0.22 & 2.03(31) &  5.5  & -- & -- & -- & -- & 45.2360(55) & 0.22 &  5.36(23) &   4.8\\
$\nu_3+6f_{orb}$ & 46.1846(70) & 0.21 & 2.02(32) &  4.9  & -- & -- & -- & --  & -- & -- & -- & --\\
$\nu_3+7f_{orb}$ & 47.1335(91) & 0.16 & 2.16(42) &  3.7  & -- & -- & -- & --  & -- & -- & -- & --\\ [0.5ex]
$\nu_4$          & 43.3803(68) & 0.22 & 4.89(31) & 5.3 & 43.3754(44) & 0.29 & 1.51(19) & 4.4 & 43.3649(35) & 0.34 &  2.78(15) &   7.9\\
\hline 
\end{tabular}
\normalsize

{Notes: The formal errors on the amplitudes are $\pm$ 0.07 mmag in Sector 34, $\pm$ 0.06 mmag in Sector 61, and $\pm$ 0.05 mmag in Sector 88, respectively. Formal errors on the frequencies and phases (in units of the last two significant digits) are listed in parentheses after the actual values. Phases are calculated with respect to the central primary eclipse at BJD 2459229.2182. The detection threshold for independent signals in the Sector 34 data is $SNR>4.8$, whereas $SNR>5.0$ was used for the Sector 61 and 88 data. Pulsation mode $\nu_2$ is not detected ($SNR < 1.5$ in all cases) and listed in italics for emphasis.}
\label{elcmi_frqs}
\end{table*}

The \'Echelle Diagram derived this way is fairly simple, but shows a faint, extended ridge around mode $\nu_3$ (Fig.\,\ref{elcmiech}, upper panel). This is a sign of spatial filtering (eclipse mapping), a known phenomenon amongst pulsating stars in eclipsing binaries. In brief, when part of the surface of a nonradially pulsating star is covered during an eclipse, only the light variations on the visible part of the star can be measured, modifying the observed pulsation amplitudes and phases depending on the type of pulsations modes excited \citep[e.g.,][and references therein]{2023A&A...671A.121V, 2025A&A...693A.259B}. 

\begin{figure}
\includegraphics[width=\linewidth,viewport=35 70 385 460]{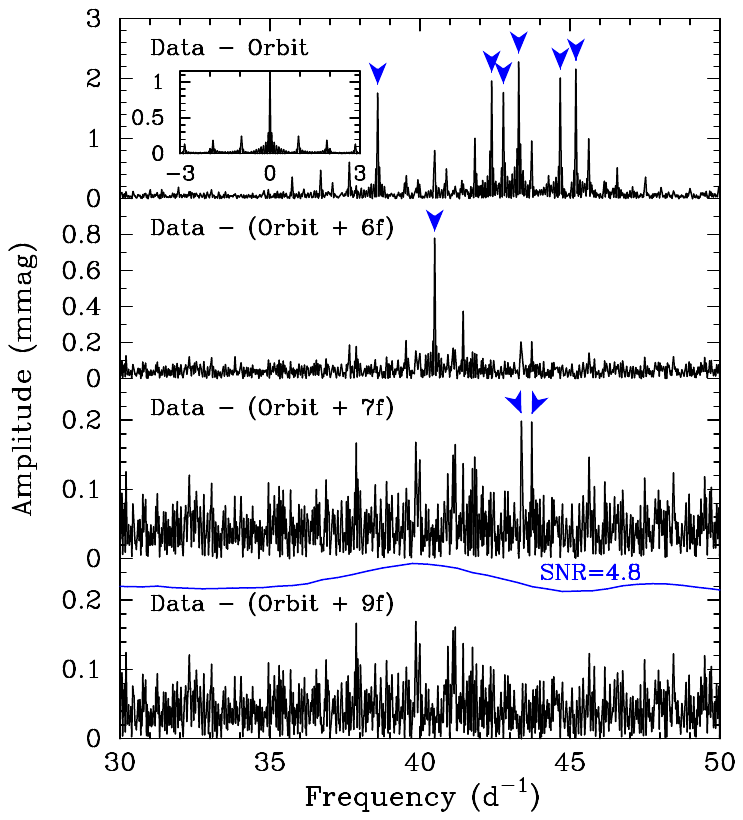}
\caption{Same as Fig.\,\ref{elcmi_fts}, but for the data after the primary eclipses have been removed. The inset in the top panel is the spectral window function of this data set.}
\label{elcmi_wfts}
\end{figure}

The effect of spatial filtering occurs when the pulsating star is attenuated by its companion; when the pulsator blocks some light of the constant star only the background light changes in strength. For this reason the frequency analysis was repeated using light curves once with the primary, and once with the secondary eclipses removed. As there was no change in the derived set of pulsation frequencies in the latter case, the hotter component of the binary is identified as the pulsator. Figure\,\ref{elcmi_wfts} shows the frequency analysis after the primary eclipses have been removed from the Sector 34 data. The list of pulsation-related frequencies derived from all sectors (with sufficient cadence) with the primary eclipses removed is given in Table\,\ref{elcmi_wprifrqs}, and the corresponding \'Echelle Diagram can be found in Fig. \ref{elcmiech} (lower panel). The results from each individual sector are highly consistent, keeping in mind that the pulsation amplitudes are somewhat variable over time, precluding the detection of some of the weaker signals in Sectors 61 and 88, respectively.

\begin{table*}
\centering
\caption{Same as Table 1, but from the data after the primary eclipse has been removed.}
\scriptsize
\begin{tabular}{l|cccr|cccr|cccr} 
\hline
\hline
ID & Frequency & Ampl. & Phase & S/N & Frequency & Ampl. & Phase & S/N & Frequency & Ampl. & Phase & S/N\\ 
& (\cd) & (mmag) & (rad) & ~ & (\cd) & (mmag) & (rad) & ~  & (\cd) & (mmag) & (rad) & \\
\hline
 & \multicolumn{4}{c}{Sector 34} &  \multicolumn{4}{c}{Sector 61} &  \multicolumn{4}{c}{Sector 88}\\
\hline
\hline
$\nu_1-f_{orb}$ & 42.7776(06) & 2.31 &  2.47(03) & 49.1 & 42.7779(07) & 1.88 &  3.76(03)	 &  30.8 & 42.7784(11) & 1.12 &  3.65(05) & 23.2\\
$\nu_1$         & 43.7265(56) & 0.23 &  2.79(25) &  5.1 & {\it 43.7268} & -- & -- & -- & 43.7273(66) & 0.19 &  3.34(28) &  4.1 \\ 
$\nu_1+f_{orb}$ & 44.6754(05) & 2.48 &  2.44(02) & 56.6 & 44.6757(06) & 2.30 &  3.78(03)	 &  35.9 & 44.6762(11) & 1.22 &  3.60(04) & 27.2\\ [0.5ex]
$\nu_2-f_{orb}$ & 43.2950(07) & 1.96 &  5.15(03) & 40.9 & 43.2953(04) & 3.04 &  4.09(02)	 &  48.0 & 43.2954(08) & 1.71 &  5.42(03) & 35.8 \\
$\nu_2$         & {\it 44.2438} & -- & -- & -- & {\it 44.2441} & -- & -- & -- & {\it 44.2442} & -- & -- & --\\
$\nu_2+f_{orb}$ & 45.1928(07) & 1.77 &  1.96(03) & 40.6 & 45.1931(04) & 2.89 &  0.93(02)	 &  45.6 & 45.1932(08) & 1.65 &  2.18(03) & 36.9 \\ [0.5ex]
$\nu_3-2f_{orb}$ & 38.5930(07) & 1.86 & 5.12(03) & 36.7 & 38.5918(10) & 1.24 &  0.12(05)	 &  20.4 & 38.5935(09) & 1.50 &  0.12(04) & 25.0 \\
$\nu_3$         & 40.4908(15) & 0.84 &  5.22(07) & 15.9 & 40.4896(23) & 0.56 &  0.09(10)	 &   9.1 & 40.4913(20) & 0.66 &  0.21(08) & 11.3\\
$\nu_3+2f_{orb}$ & 42.3887(08) & 1.70 & 2.00(03) & 34.7 & 42.3874(11) & 1.15 &  3.31(05)	 &  18.6 & 42.3891(10) & 1.28 &  3.20(04) & 25.2 \\ [0.5ex]
$\nu_4$         & 43.3798(65) & 0.20 &  4.75(30) &  4.2 & 43.3657(46) & 0.28 &  6.20(21) & 4.4 & 43.3712(46) & 0.28 &  4.35(20) &  5.9 \\
\hline 
\end{tabular}
\normalsize

{Notes: The formal errors on the amplitudes are $\pm$ 0.06 mmag in all sectors of data.}
\label{elcmi_wprifrqs}
\end{table*}

\begin{figure*}
\includegraphics[width=\textwidth,viewport=60 70 568 350]{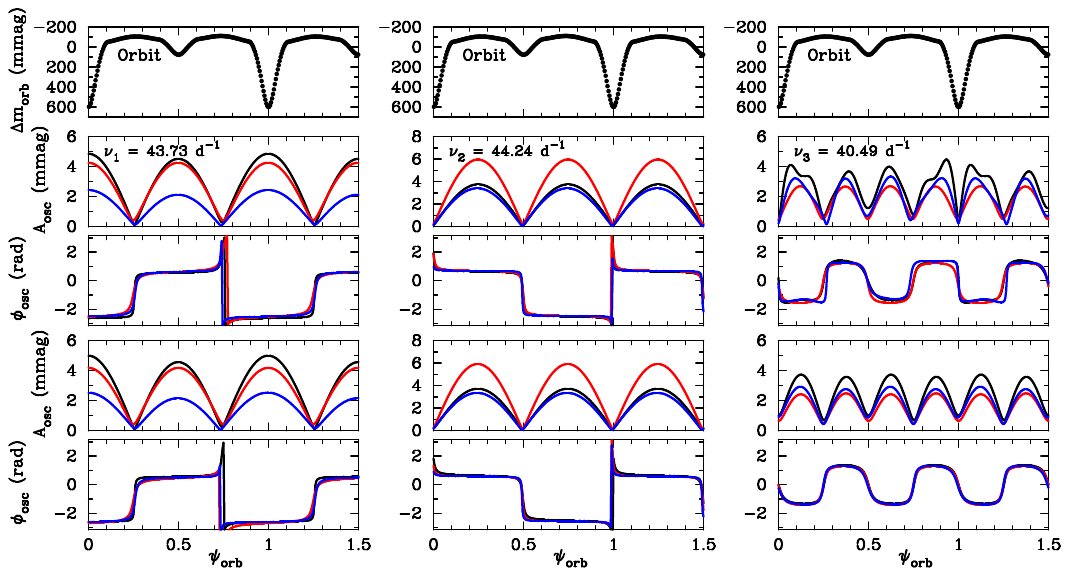}
\caption{The run of the pulsation amplitude and phase for the three main modes of EL CMi. Black: Sector 34 data. Red: Sector 61 data. Blue: Sector 88 data. Upper panel: orbital light curve. Second and third panels: using all data. Lower two panels: reconstructions from the data omitting primary eclipses. The frequencies given are for the multiplet centroids.}
\label{elcmiamppha}
\end{figure*}

With that information in hand, the runs of pulsation amplitude and phase for the three dominant modes were reconstructed using the method described by \cite{2022ApJ...928L..14J} (their Sect. 3.2). Figure \ref{elcmiamppha} shows the results.

Whereas the amplitudes of the three pulsation modes varied from one sector of data to the next, the runs of the pulsational phases over the orbit are essentially identical. The higher amplitude of $\nu_3$ and lower noise level in Sector 34 make the effects of spatial filtering during primary eclipse detectable, whereas there are only indications of it in the Sector 88 data, and they are lost in the observational noise in Sector 61. To identify the pulsation modes by comparison with the pulsation amplitude and phase curves theoretically predicted by \cite{2025ApJ...979...80F}, it is more convenient to use the reconstructions from the data with the primary eclipses removed.

The amplitude and phase modulation of mode $\nu_1$ over the orbit is consistent with that of a $Y_{10x}$ dipole mode, whereas the run of mode $\nu_2$ is that of a $Y_{10y}$ dipole mode (see \cite{2025ApJ...979...80F} for visualizations of these modes). Such modes have already been previously reported \citep{2024MNRAS.528.3378Z,2024ApJ...975..121J}. The behavior of the pulsation amplitude and phase depending on orbital phase is clearly different for mode $\nu_3$; it is that of a quadrupole $Y_{22-}$ mode, which is thus observationally detected for the first time.

As discussed by \cite{2025ApJ...979...80F}, the $Y_{22-}$ mode is a superposition of $Y_{2+2z}$ and $Y_{2-2z}$ modes. This behaves like an $l = m = 2$ pattern that is wrapped around the z-axis, with maxima/minima offset from the tidal axis by an angle of $\pi/4$, and it is a standing mode that does not propagate around the z-axis.

\subsection{Spectroscopy}
\label{sec:spec}
Spectroscopic observations of EL CMi were carried out with the \texttt{ALFOSC} instrument in spectroscopy mode at the 2.56-m Nordic Optical Telescope (NOT) on Observatorio del Roque de los Muchachos, La Palma, Spain during the nights from 2025 January 23/24 to 27/28. Grism \#18 was used to result in a wavelength coverage of 3450--5350\,\AA ~with a spectral resolution of 2000 using a $0.5''$ slit. Integration times of 300 s were used to result in a SNR of 65 -- 145 around 4500\,\AA ~depending on weather conditions. Each observation consisted of two or three consecutive integrations.

The data were reduced with the \texttt{PypeIt} package \citep{pypeit:joss_pub, pypeit:zenodo}, using standard day-time calibrations as well as arc frames taken immediately after the on-target integrations in order to avoid spurious variability due to flexure, and applying barycentric corrections.

\begin{figure}
\includegraphics[width=\linewidth,viewport=35 30 390 280]{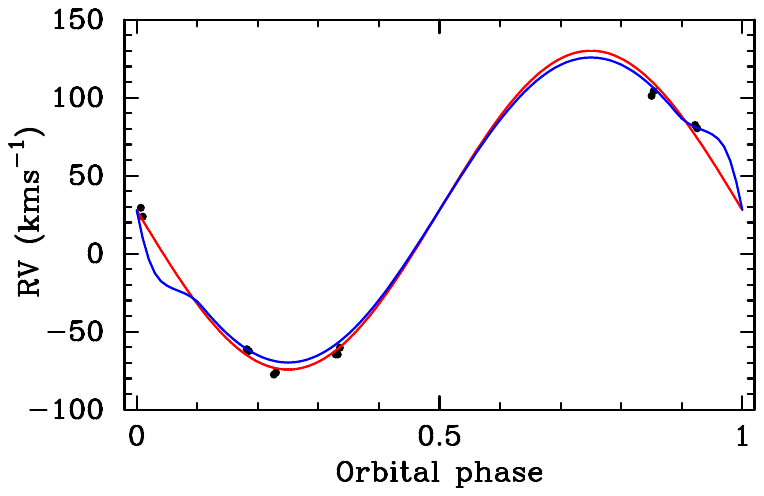}
\caption{The orbital \texttt{NOT} radial velocities (black points) with the simple Keplerian (red line) and the \textsc{PHOEBE} (blue line) fits that also include the predicted Rossiter-McLaughlin signature \citep{1924ApJ....60...15R,1924ApJ....60...22M} The uncertainties on the measurements are about the size of the plot symbols.}
\label{elcmirv}
\end{figure}

Radial velocities were extracted from these data using the \texttt{IRAF} fxcor task. To facilitate the cross-correlation a synthetic spectrum with \teff=7750\,K, \logg=4.0, \vsini=90\,\kms was computed with the {\tt SPECTRUM} code \citep{1994AJ....107..742G}. The radial velocities so obtained are listed in Table\,\ref{tab:rvs}. Using the ephemeris in Sect.\,\ref{sec:phot} and assuming a circular orbit yields a radial velocity amplitude of $K_1=102.3\pm3.0$\,\kms for the primary star and a systemic velocity $\gamma=+28.1\pm3.0$\,\kms. The stated uncertainties include both the formal error of the fit and the spectrograph's wavelength calibration error. Figure\,\ref{elcmirv} shows the radial velocity measurements in comparison to this and a model fit to be derived in the next section.

\begin{table}
\centering
\caption{Radial velocity (RV) data for the primary component of EL CMi.}
\begin{tabular}{ccc} 
\hline
HJD -- 2460000 & $\phi_{\rm orb}$ & $RV$\\
(d) & & \kms \\
\hline
 699.68559 &  0.3292 & $-$64.6\\
 699.68918 &  0.3326 & $-$64.6\\
 699.69278 &  0.3360 & $-$60.3\\
 700.40028 &  0.0073 & +29.4\\
 700.40387 &  0.0107 & +23.8\\
 700.63187 &  0.2271 & $-$77.5\\
 700.63546 &  0.2305 & $-$76.2\\
 701.63853 &  0.1823 & $-$61.3\\
 701.64213 &  0.1857 & $-$62.4\\
 702.41848 &  0.9224 & +82.5\\
 702.42207 &  0.9258 & +80.4\\
 703.39660 &  0.8506 & +101.1\\
 703.40020 &  0.8540 & +104.2\\
\hline 
\end{tabular}

{Notes. The error of each individual radial velocity measurement is some $\pm 5$\,\kms.}
\label{tab:rvs}
\end{table}

Finally, the individual spectra were corrected for radial velocity shifts, co-added and compared to model spectra of different temperatures and surface gravities. The best fit was found for \teff=7750\,K, \logg=4.0, \vsini=\,90\,\kms (hence the previous choice of template). The primary star appears to be somewhat underabundant in heavy elements compared to the Sun ($[M/H]=-0.3\pm0.2$). However, our spectra are insufficient to tell whether this is a true underabundance or whether the primary star has some mild chemical peculiarity of the $\lambda$ Bootis type.

\section{System Parameters}
\label{sec:system_parms}

\subsection{Data Availability}

The Gaia DR3 catalog \citep{gaia_mission,gaia_dr3} gives a $G$ magnitude of 11.82 and $(B_P-R_P)=0.34$, which correspond to the mean values out of eclipse. The Gaia DR3 parallax of the system is $0.977\pm0.019$\,mas. The reddening maps of \cite{2019ApJ...887...93G} imply $E(g-r)=0.025\pm0.015$, consistent with \cite{1998ApJ...500..525S} ($A_g=0.139$) which is taken as an upper limit to the interstellar reddening of the system given its Galactic latitude of $+15\deg$.

With the reddening relations by \cite{2023ApJS..264...14Z} and the transformations from the Gaia photometric system given in Table 5.9 of the Gaia DR3 release notes, $E(B_P-R_P)=0.030\pm0.018$ and $V=11.86$ was derived. The standard relations by \cite{2013ApJS..208....9P} then suggest \teff=7670\,K, in very good agreement with the spectroscopic estimate above. With a bolometric correction of $BC=+0.015$, $A_v=0.08$ and the Gaia parallax $M_{bol}=+1.74\pm0.07$ is derived (with the error budget dominated by the uncertainties in interstellar extinction and the $V$ magnitude). With the bolometric magnitude of the Sun \citep{2010AJ....140.1158T}, $M_{bol,\sun}=4.74$, a mean out of eclipse system luminosity of $L=15.8\pm1.1$\,L$_{\sun}$ results.

Archival spectral flux measurements from 0.15 to 11.6 microns were collected. These data were taken from VizieR SED\footnote{A.-C. Simon \& T. Boch: \url{http://vizier.cds.unistra.fr/vizier/sed/}} \citep{2000A&AS..143...23O} which, in turn, utilizes systematic sky coverage of such surveys such as Pan-STARRS \citep{ps1_survey}, SDSS \citep{1998AJ....116.3040G}, 2MASS \citep{2mass_survey}, WISE \citep{wise}, and Galex \citep{galex_catalog}. For EL CMi we found 26 spectral points that we were able to use.

\subsection{General Approach}

In what follows the EL CMi system parameters are derived using two different approaches. In the first, the \textsc{phoebe2} software is utilized to analyze the eclipsing light curve obtained with {\it TESS} plus the radial velocity data from the primary star. In the second approach our own spectral energy distribution (SED) fitting code coupled with a relatively simple eclipsing light curve emulator is used to jointly analyze the {\it TESS} light curve, the available SED data, and the radial velocities for this source. Each approach is detailed in the following subsections.

\subsection{PHOEBE Analysis}
\label{sec:phoebe}

The new-generation Wilson-Devinney code \textsc{phoebe2} \citep{2016ApJS..227...29P} was used to analyze the {\it TESS} eclipsing light curve of EL CMi and to simultaneously take into account the measured radial velocity data points of the primary star. A Monte Carlo Markov Chain (MCMC, \citealt{2013PASP..125..306F}) wrapper was utilized to evaluate the uncertainties in the fitted parameters by sampling over a broad-range of uninformed priors on the ratio of effective temperatures, primary radius, primary mass, mass ratio, secondary radius, orbital inclination and systemic velocity. 

\begin{figure}
\includegraphics[width=\linewidth,viewport=40 40 395 334]{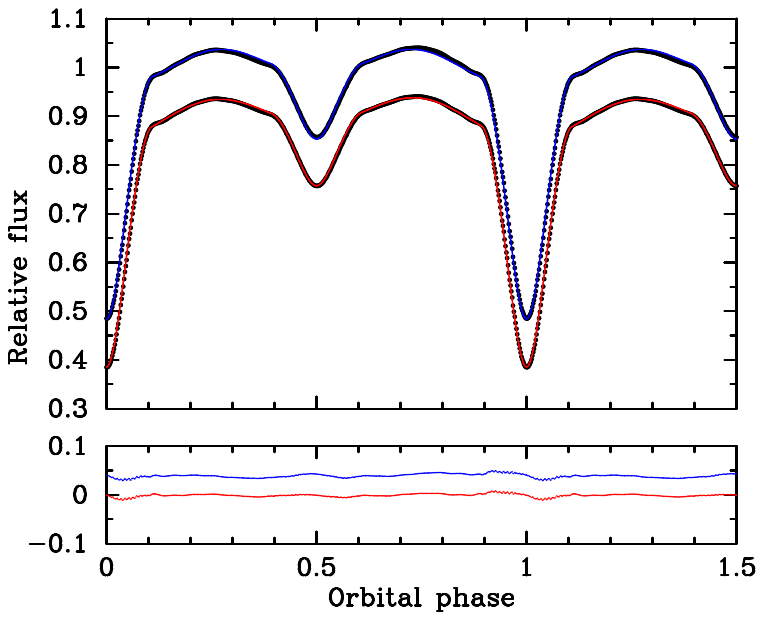}
\caption{Upper panel: the fit to the phase binned {\it TESS} PDC light curve from Sector 34 (black points) compared to the fit derived with \textsc{phoebe2} (blue line) and the combined SED + light curve fitting approach (red line, data and fit shifted by 10\% in intensity for clarity). Lower panel: the residuals from these fits. Blue: \textsc{phoebe2}, red: SED + light curve fitting, shifted by 4\% in intensity for clarity.}
\label{elcmilcfit}
\end{figure}

The fitted light curve is shown in Fig.\,\ref{elcmilcfit}. The best fitting parameters that result from this fit are shown in the first results column of Table \ref{tbl:parms1}. Note in particular that we simply fixed the $T_{\rm eff}$ of the primary at 7750 K, as deduced from the spectroscopy (see Sect.~\ref{sec:spec}). Both stars were modeled with \citet{2003IAUS..210P.A20C} model atmospheres (where these atmospheres were used to derive the emergent intensity of each stellar surface element as well as the limb darkening via \textsc{PHOEBE2}'s interpolated limb darkening scheme, and with default gravity brightening ($\beta$=0.32) and bolometric albedos ($\alpha$=0.6).

The mass and $T_{\rm eff}$ of the pulsating $\delta$~Scuti star are typical for such pulsators. On the other hand, the secondary, with a mass of only 0.93 M$_\odot$ and radius of nearly 2 R$_\odot$, strongly hints at a star which has lost its envelope during a prior episode of mass transfer. This is discussed further in the next section.

\subsection{Light curve plus SED Analysis}
\label{sec:LC+SED}

The SED of a binary can yield important information on the two stellar radii and the two $T_{\rm eff}$ values, provided that the distance to the source is known accurately. However, the SED by itself is not generally sufficient to yield all four of these parameters. Additionally, the eclipsing light curve contains key information on such quantities as the ratio of the two $T_{\rm eff}$ values, and the ratios of the stellar radii to the semi-major axis. But, the latter requires the system mass in order to translate $R_1/a$ and $R_2/a$ to information about the physical stellar radii. In turn, radial velocities can provide information about the masses. In any case, supplemental information about the radii and effective temperatures can indeed facilitate the inference of a number of the stellar properties from the SED data.

Under the assumption that the two stars in the binary have evolved coevally (i.e., with no prior episode of mass transfer), the SED can be fit with just four basic parameters: $M_1$, $M_2$, system age, and an interstellar extinction, $A_V$. If the metallicity is not known from the spectral observations, it can be added as a fifth parameter.

The mass and age yield the radius and $T_{\rm eff}$ of the two stars via stellar evolution tracks, and the SED can then be fit, provided that there is some other supplemental information such as ratios of radii or $T_{\rm eff}$ from a light curve analysis, e.g., from \textsc{phoebe2}. Since \textsc{phoebe2} does not have a built-in SED fitter, the light curve fitting with \textsc{phoebe2} and the SED fitting could be iterated with separate codes. However, it was chosen instead to write a combined SED fitter with a simplified light curve emulator that can be used to jointly fit for both the SED and the eclipsing light curve. The code follows the description by \cite{2024ApJ...975..121J}. Briefly, the simplified light curve emulator utilizes (i) spherical stars with limb darkening, and (ii) sets of sines and cosines to represent the out-of-eclipse behavior, including ellipsoidal light variations, illumination effects, and Doppler boosting \citep[see, e.g.][]{2011ApJ...728..139C}. 

The SED portion of the fitting code utilizes MESA Isochrones and Stellar Tracks (MIST) evolution tracks (\citealt{mist_ii}; \citealt{mist_i}; \citealt{Paxton2011}; \citealt{Paxton2015}; \citealt{Paxton2019}). For any given mass, age, and metallicity, the tracks provide the stellar radius and $T_{\rm eff}$. The stellar atmosphere models were taken from \citet{castelli_kurucz}. The SED fitting portion of the code has been extensively tested on a number of multistellar systems, including binaries \citep{2022ApJ...936..123J},
triples \citep{2022MNRAS.513.4341R}, and even a sextuple star system \citep{2021AJ....161..162P}. For EL CMi, metallicities of $[Fe/H] = 0.0$ and $-0.3$ were tested in keeping with what was learned from the stellar spectra (see Sect.~\ref{sec:spec}). There was not much difference in the results depending on the metallicity. 

This code was used to jointly fit the light curve, SED, and radial velocity data for EL CMi, in agreement with the \textsc{phoebe2} result. It was found that there are no viable solutions where the two stars evolve in a coeval manner. Thus, there must have been a prior episode of mass transfer onto the current primary star (the pulsator). Therefore the fitted parameters were changed to the following 6 quantities: primary and secondary mass ($M_1$ and $M_2$), radius and $T_{\rm eff}$ of the secondary star ($R_2$ and $T_{\rm eff,2}$), the age of the primary star (since its rejuvenating accretion episode, $\tau$), and $A_V$. 

\begin{figure}
\includegraphics[width=\linewidth,viewport=10 5 452 303]{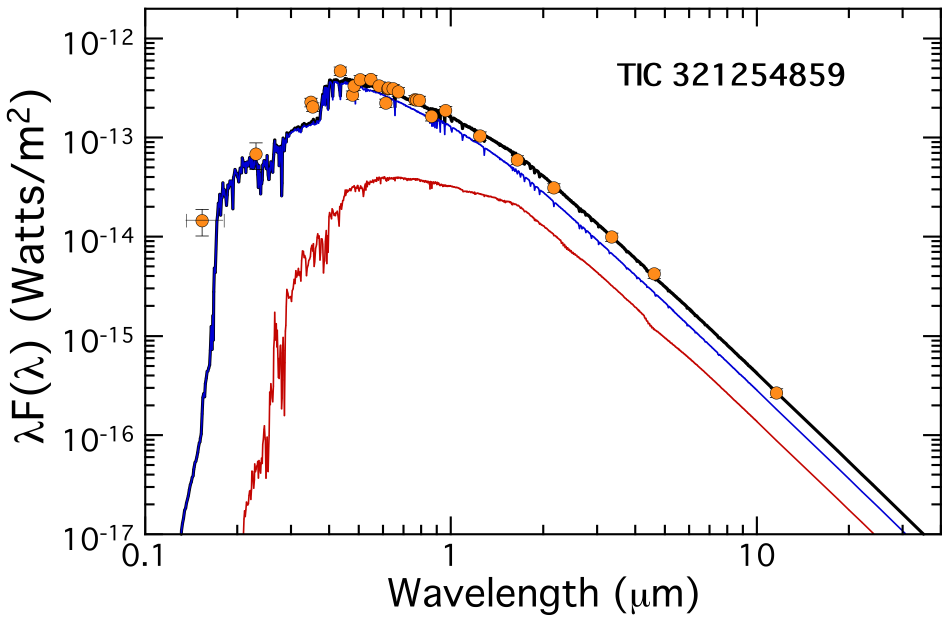}
\caption{The fit to the SED of the EL CMi system. The orange dots with the error bars are the observational values. The blue line is the SED of the primary star, the red line the SED that of the secondary and the black line is the sum of the two.}
\label{elcmiSED}
\end{figure}

The fitted light curve is shown in Fig.\,\ref{elcmilcfit}. Figure\,\ref{elcmiSED} shows a comparison of the observed to the theoretical SED derived above as well as the fitted light curve. The agreement is satisfactory. Only the point in the far ultraviolet is somewhat off. However, this may be due to the wide bandwidth of the \texttt{GALEX} FUV filter and the steep continuum slope in this wavelength range \citep[see][]{2024ApJ...975..121J}.

The binary system parameters are given in two columns of Table \ref{tbl:parms1} along with some other derived parameters in Table \ref{tbl:parms2}. The second column of the results parameters in Table \ref{tbl:parms1} is for the case where there are no prior restrictions placed on the radius of the secondary star, except that it cannot overfill its Roche lobe. The third results column in Table \ref{tbl:parms1} is for the restricted case where the secondary star always just fills its Roche lobe. This is motivated by evolutionary studies (see Sect.\ref{sec:evolution}) which indicate that after the episode of rapid mass transfer, the donor star will remain close to filling its Roche lobe for 1.3 Myr after the bulk of the mass transfer has ceased.

A comparison of the three results columns in Table \ref{tbl:parms1} paints a consistent picture of the stellar parameters for EL CMi. The main difference in the three results columns arises in the radii, depending on whether the secondary is assumed to fill its Roche lobe or not. In the former case, the primary is $24\% \pm 11\%$ larger than the secondary, while in the latter case, the two stars are more nearly equal in size, with the secondary being slightly larger. 

\begin{table*}
\centering
\caption{Binary Parameters of TIC 321254859 (EL CMi) with two different methods. }
\begin{tabular}{cccc}
\hline
\hline
Parameter & \textsc{PHOEBE} & Light curve + SED (1) & Light curve + SED (2) \\
\hline
   $M_1$ [M$_\odot$] & $1.78\pm0.03$ & $1.70\pm0.03$ & $1.75 \pm 0.03$  \\
   $M_2$ [M$_\odot$] & $0.93 \pm 0.01$ & $0.96 \pm 0.03$ & $0.97 \pm 0.03$ \\
   $R_1$ [R$_\odot$]  & $1.88\pm0.01$  & $1.85\pm0.11$ & $2.14\pm0.11$  \\
   $R_2$ [R$_\odot$]  & $1.97\pm0.01$  & $2.00 \pm 0.03$ & $1.72\pm0.12$ \\
   $T_{\rm eff,1}$ [K] & $7750$  (fixed) & $7779\pm120$ & $7566\pm120$  \\
   $T_{\rm eff,2}$ [K] & $5040\pm10$  & $5216\pm123$ & $5113\pm120$  \\
   $i^{\circ}$ [deg] &$78.6 \pm 0.05$& $76.9 \pm 0.7$ & $80.6 \pm 2.3$\\
   \hline
   $L_1$ [L$_\odot$] &$11.5 \pm 0.6$ & $11.3\pm1.2$ & $13.6\pm1.2$  \\
   $L_2$ [L$_\odot$] & $2.26\pm0.05$ & $2.7\pm0.27$ & $1.80\pm3.2$  \\
   $q$  & $0.523\pm0.004$ & $0.566\pm0.019$  & $0.554\pm0.019$   \\
   $\gamma$ [km s$^{-1}$] & $28.0 \pm 1.4$  & {\rm n/a} & {\rm n/a}  \\
 \hline
\label{tbl:parms1}
\end{tabular}

{Notes. All the fits make use of the radial velocity data. (1) Light curve + SED joint fit is done under the assumption that the secondary star fills its Roche lobe. (2) Light curve + SED joint fitting done with the assumption that the secondary does not overfill its Roche lobe. }
\end{table*}

\begin{table*}
\centering
\caption{Other parameters of EL CMi from simultaneous
light curve and SED fitting. }
\begin{tabular}{ccc}
\hline
\hline
Parameter & Light curve + SED (1) & Light curve + SED (2)  \\
\hline
  $A_V$$^a$ [mag] & $0.079 \pm 0.004$ & $0.080 \pm 0.004$ \\
   $a$ & $6.04\pm0.04$ & $6.07 \pm 0.04$  \\
   $R_1/R_{L,1}$ & $0.71\pm0.04$ & $0.82\pm0.04$  \\
   $R_2/R_{L,2}$ & 1.00 (fixed) & $0.86\pm0.06$  \\
    $R_1/a$  & $0.305\pm0.017$ & $0.352\pm0.017$  \\
   $R_2/a$   & $0.331\pm0.003$ & $0.282\pm0.019$  \\
   $K_1$$^b$ [K] &  $102.2\pm2.5$ & $102.6\pm2.5$  \\
   $K_2$$^c$ [K] &  $180.5\pm2.8$ & $185.4\pm2.4$  \\
   $\cos(\omega t)$$^d$ [ppt]  & $-16.3 \pm 0.3$ & $-16.3 \pm 0.3$\\
   $\cos(2\omega t)$$^d$ [ppt]  & $-37.2 \pm 0.2$ & $-37.2 \pm 0.2$\\
   $\sin(\omega t)$$^d$ [ppt]  & $-2.49 \pm 0.08$ & $-2.49 \pm 0.08$\\
 \hline
\label{tbl:parms2}
\end{tabular}

{Notes. Fits were done in the same way as indicated in the caption to Table \ref{tbl:parms1}. (a) The prior on $A_V$ was set to cover the limited range of 0.07 to 0.09. (b) This result is essentially just the prior on $K_1$. (c) This is the prediction of the $K$ velocity of the secondary based on the fitted system parameters. (d) These represent the primary illumination term, the primary ellipsoidal light variations term, and the Doppler boosting term, respectively. Because this latter measured value is an order of magnitude larger than is physically plausible, and of the opposite sign, this is attributed to star spots on the relatively cool companion star.}
\end{table*}

\section{System evolution}
\label{sec:evolution}

As the data analysis suggests that the secondary has a radius too large to be explicable as a product of a single-star evolution, it is logical to suspect that EL CMi is a product of a system that has passed through a mass transfer phase.
To model the system as the result of a prior episode of mass transfer, the \textsc{mesa} code \citep[Modules for Experiments in Stellar Astrophysics,][version 23.05.1]{Paxton2011, Paxton2013, Paxton2015, Paxton2018, Paxton2019, Jermyn2023} was used with the \textsc{mesa-binary} module under the implicit mass transfer rate assumption.

To find a model that fits the parameters of the system, in particular the masses, radii, effective temperatures and the orbital period, a grid of binary evolution tracks for initial periods spanning a range from 0.6 to 10\,d and for initial masses in the range of 1.0 to 3.0\,\Msun, with a constant step size of 0.2\,d and 0.2\Msun\ respectively, was built. 
Additionally, the results from Section \ref{sec:system_parms} were used and, accounting for a small mass loss, the constraints that the initial total mass of the system should not be less than 3.0 \Msun, but no more than 3.5 \Msun\ were imposed. After obtaining an initial fit, the grid resolution was refined around the best-matching model by reducing the steps to 0.02\,\Msun\ in mass and 0.05\,d in orbital period.

Four different values of metallicity, $Z=0.005$, 0.010, 0.014 and 0.020, corresponding to $\mathrm{[M/H]} = -0.47, -0.16, -6.5 \times 10^{-3}$ and 0.16 respectively, were used and scaled using the linear scaling relations by \cite{Choi2016} to determine the initial hydrogen and helium abundances.
The AGSS09 \citep{Asplund2009} initial chemical composition of the stellar matter and the OPAL opacity tables supplemented with data from \cite{Ferguson2005} for lower temperatures were used. Convective instability was treated using the Ledoux criterion, combined with the mixing length theory based on the \cite{Henyey1965} model. In regions stable according to the Ledoux criterion but unstable by the Schwarzschild criterion, semi-convective mixing using a scaling factor of $\alpha_{\rm sc} = 0.1$ was used, following the formalism of \cite{Langer1985}. To mix regions with mean molecular weight inversion, such as those formed during mass accretion, thermohaline mixing using the formalism of \cite{Kippenhahn1980} was applied, with an $\alpha_{\rm th} = 1$ coefficient. As these models include neither rotation nor rotationally induced mixing, a minimum diffusive mixing coefficient of $D = 10\  \rm cm^2\,s^{-1}$ was used that also permits to smooth out numerical noise or discontinuities in internal profiles. 

The values of free stellar structure parameters were fixed according to the following theoretical models and asteroseismic calibrations. Given the primary’s mass, the convective core overshooting parameter was set to $f_{\rm ov} = 0.02$, in agreement with the values predicted by \citet{Claret2016}, for both components. Furthermore, the mixing length parameter and the fraction of mass lost from the system during mass transfer were set to $\alpha_{\rm MLT} = 1.5$ and $\beta = 0.2$, respectively, in agreement with asteroseismic determinations for $\delta$ Scuti stars in post-mass-transfer binaries \citep[e.g.][]{2022MNRAS.514..622M}.

\begin{figure}
\includegraphics[width=1.0\linewidth]{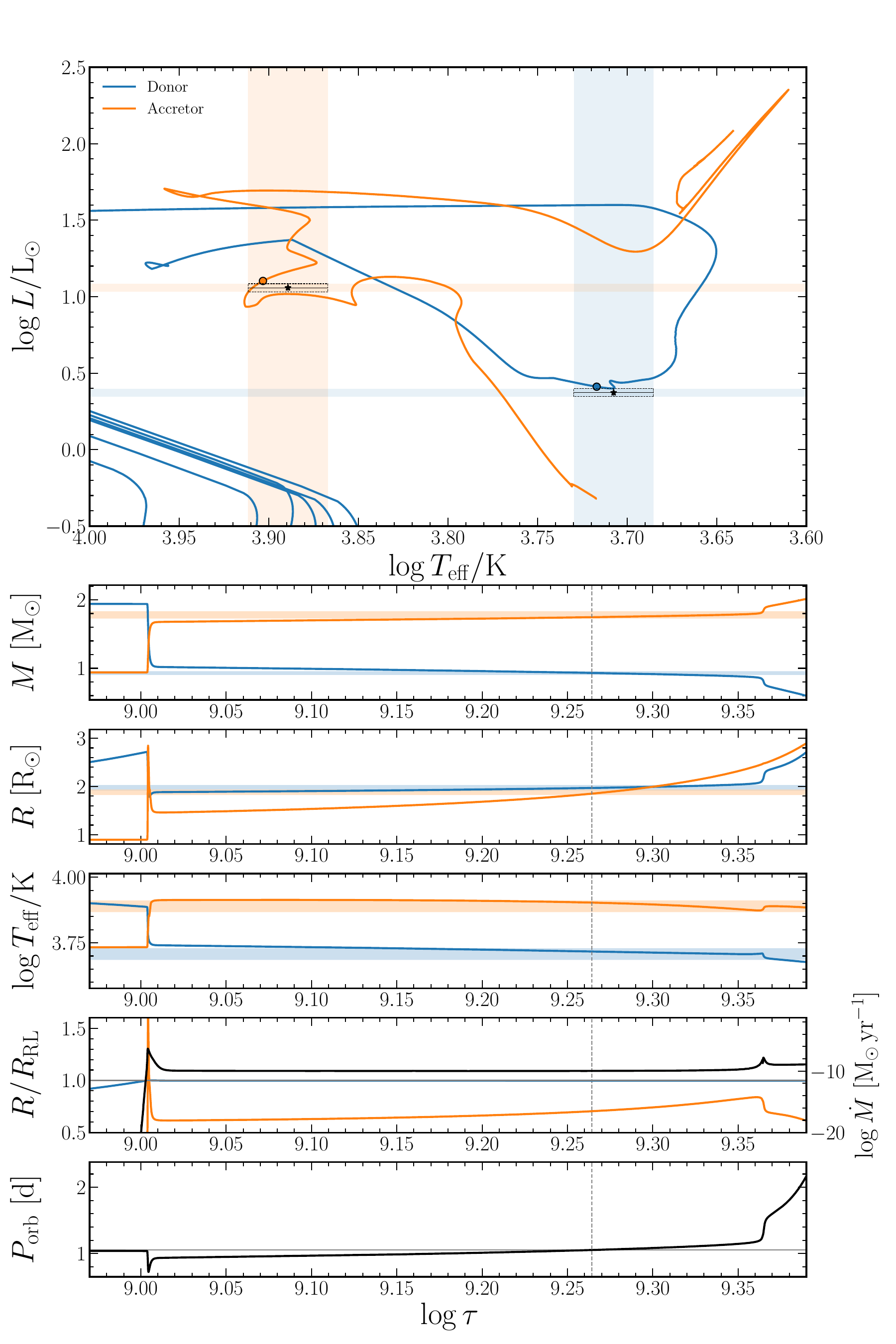}
\caption{The HR Diagram and the time evolution of the stellar masses, radii, effective temperatures, Roche lobe filling factor and the orbital period for the binary evolution track that best reproduces the observed parameters of the system. This track was computed for the following initial parameters: orbital period $P_{\rm orb,0} = 1.05$\,d, primary and secondary masses $M_{\rm don,0} = 1.94$\,\Msun\ and $M_{\rm acc,0} = 0.94$\,\Msun, metallicity $Z = 0.014$, overshooting parameter $f_{\rm ov} = 0.02$, mixing-length parameter $\alpha_{\rm MLT} = 1.5$, and mass transfer efficiency $\beta = 0.2$. The evolutionary tracks of the accretor and the donor are shown with orange and blue lines, respectively, and the same colors are used to indicate the observed ranges of the corresponding parameters, which are marked with shaded bands on the plots. The model at the time corresponding to the observed orbital period is marked with vertical dashed lines in the lower panels and with circular markers on the HR diagram. }
\label{MESAfit}
\end{figure}

From each set of binary evolutionary tracks, models were selected that agree with the measured current-epoch orbital period, and for those models a normalized $\chi^2$ discriminant using observed and model masses, radii and effective temperatures was calculated. This resulted in a set of 40 models that fit the observations within 1$\sigma$ and cluster around $P_{\rm orb,0} \in (1.0, 1.05)$\,d, $M_{\rm don,0} \in (1.9, 2.06)$\,\Msun\ and $M_{\rm acc,0} \in (0.9, 1.0)$\,\Msun\ for metallicities $Z=0.014$ and $Z=0.02$. A visualization of the best-fitting model is presented in Figure\,\ref{MESAfit}. In the upper panel the evolution tracks are shown in the Hertzsprung-Russell diagram, while the lower panels depict the time evolution of the stellar masses, radii, effective temperatures, Roche lobe filling factor, and the orbital period.
The tracks shown on the plots were computed for the following initial parameters: orbital period $P_{\rm orb,0} = 1.05$\,d, primary and secondary masses $M_{\rm don,0} = 1.94$\,\Msun\ and $M_{\rm acc,0} = 0.94$\,\Msun, metallicity $Z = 0.014$, overshooting parameter $f_{\rm ov} = 0.02$, mixing-length parameter $\alpha_{\rm MLT} = 1.5$, and mass transfer efficiency $\beta = 0.2$.

According to these models, the system is approximately $1.7 \pm 0.3$ Gyr old and is still undergoing mass transfer. However, the mass transfer rate is currently fairly low at $\dot{M} \approx 10^{-10} \rm\ M_{\odot}\,yr^{-1}$$ $. The most rapid phase of mass transfer began when the system was $\sim 1$ Gyr old, and peaked at $\dot{M} \approx 10^{-6} \rm\ M_{\odot}\,yr^{-1}$. At that time, the donor star was in the later part of its main sequence evolution, with a central hydrogen abundance of $X_c = 0.26$, while the accretor had barely evolved from the zero-age main sequence. That suggests that the binary went through type A mass transfer (MT) \citep{Kippenhahn1967}. Currently, after rejuvenation of the accretor and mass-ratio reversal, the donor is close to central hydrogen exhaustion ($X_c < 0.1$), and has retained a sufficiently thick hydrogen envelope to soon start ascending the Red Giant Branch. On the other hand, the accretor is currently evolving through the main sequence, with a central hydrogen abundance of $X_c = 0.45$. These parameters are summarized in Table\,\ref{tbl:evol_params}.

\begin{table}
\centering
\caption{Summary of the key evolutionary parameters of the binary model.}
    \begin{tabular}{lc}
    \hline
    \hline
    Parameter & Value \\
     \hline
        Systems' age $\tau$ & $1.7 \pm 0.3\,\mathrm{Gyr}$ \\
        Mass transfer rate $\dot{M}$ & $\sim 10^{-10}\,\mathrm{M_{\odot}\,yr}^{-1}$ \\
        Donor's current $X_c$ & $< 0.1$ \\
        Accretor's current $X_c$ & $0.45$ \\
        Systems' age at the beginning of MT & $1.01\,\mathrm{Gyr}$ \\
        Time since the main phase of MT & $0.65\,\mathrm{Gyr}$ \\
     \hline
    \label{tbl:evol_params} 
    \end{tabular}
\end{table}

Despite the decline in the mass transfer rate following the initially more rapid phase, the donor remains close to Roche lobe filling over an extended timescale, with $R_{\mathrm{don}} / R_{\mathrm{RL}} \lesssim 1$. Although the donor is initially driven out of thermal equilibrium, it quickly recovers it once the mass ratio reverses. From that point onward, the donor becomes the less massive star and its radius evolution, governed by hydrogen burning, begins to regulate the mass transfer rate. As a result, the subsequent mass transfer proceeds on the donor’s nuclear timescale, allowing the stable, low-rate Roche Lobe overflow configuration to persist until central hydrogen exhaustion. Over this phase, the donor maintains sufficient pressure support to avoid contraction, and the system evolves stably under slow mass exchange \citep[e.g.,][]{Kolb1990}.

Initially, the accretor temporarily departs from thermal equilibrium due to the rapid accretion, resulting in increased luminosity. However, it regains equilibrium shortly after and continues its main sequence evolution with a steadily growing mass and radius.
Once reaching the end of the mass transfer phase, the accretor will attain a mass of $2.16$\,\Msun\, cross the Hertzsprung gap and become a red giant.

As mass transfer continues, the donor gradually loses its hydrogen-rich envelope. Once mass transfer ceases, the envelope is no longer massive enough ($M=0.2$\,\Msun) to maintain efficient shell hydrogen burning or to provide the pressure required to compress the core toward helium ignition. At that point, the central temperature reaches only $\log T_c \approx 7.52$, and the helium core mass is limited to $M_{\mathrm{core}} \approx 0.17$\,\Msun, both well below the thresholds required for helium ignition. Consequently, instead of ascending the RGB up to the helium ignition, the star will evolve towards the pre-helium white dwarf (pre-He WD) phase. Despite its relatively high initial mass, the donor ultimately avoids helium ignition and becomes a low-mass helium white dwarf star.

At an age of $\sim 2.9$ Gyr, the donor star evolves into an AGB star and starts mass transfer to the helium white dwarf. At this point, the AGB star has a helium core mass of 0.488\,\Msun and a carbon/oxygen core of 0.280\,\Msun. Since the binary mass ratio is large, the mass transfer is dynamically unstable and the binary system will enter a common envelope (CE) phase. By comparing the binding energy of the envelope with the orbital energy of the system, we find that the CE cannot be ejected even if assuming a CE ejection efficiency of 1.0. Consequently, the binary system will ultimately merge into a single star.

Returning to the present evolutionary state of the system it is not clear whether the ongoing accretion affects the triaxial pulsations of the primary (the accretor), but we do not expect major effects. Since the primary is underfilling its Roche lobe and its current accretion rate is low, we do not expect substantial deviations from hydrostatic equilibrium assumed in the triaxial pulsation model. The accretion power is $L_{\rm ac} \sim G M \dot{M}/R \sim 0.003 L_\odot (\dot{M}/10^{-10} M_\odot /{\rm yr}$, which is much lower than the primary's luminosity, so its structure and pulsations are unlikely to be greatly modified.

\section{Discussion and conclusions}
\label{sec:conc}

The first objects dubbed ``tidally tilted pulsators" were explained as stars having their pulsational axis pulled into the orbital plane, and having this axis point at their close companion star at each point in the orbit. The first such pulsator, HD 74423, had a single dipole oscillation mode whose amplitude was in addition enhanced in one hemisphere, which provided motivation to coin the term ``single sided pulsator" \citep{2020NatAs...4..684H}. The second such object, CO Cam, showed multiperiodic axisymmetric pulsations enhanced in amplitude in the direction of the companion \citep{2020MNRAS.494.5118K}, whereas the third one, TIC 63328020, appeared distinctly different with its tesseral tidally tilted dipole mode \citep{2021MNRAS.503..254R}.

The idea that stars can pulsate triaxially was based on the observational studies of TIC 184743498 by \cite{2024MNRAS.528.3378Z} and of TIC 435850195 by \cite{2024ApJ...975..121J} that demonstrated that both objects exhibit a large number of tidally tilted dipole modes. However, the geometries of these binary systems, and hence the inclination of the tilted stellar pulsational axis to the line of sight made it impossible to explain those objects in terms of simple tidally tilted pulsators, as the observed amplitude and phase modulations of the individual pulsations over the orbit are predictable depending on the type of the pulsation modes and on the orbital geometry of the system \citep{2005ApJ...634..602R}. The behavior of the pulsations of these latter two stars did not conform to a simple picture of tidal tilting, but was explicable with pulsation around three different axes.

The theoretical foundation for this interpretation was provided by \cite{2025ApJ...979...80F} who showed that the different azimuthal orders of a dipole frequency multiplet are tidally coupled in those systems. Thus they align with the three principal axes of the triaxially deformed pulsating star. With this theory developed, theoretical predictions of how nonradial quadrupole oscillation modes would manifest themselves were provided. In this paper, we presented the first observational detection of such a quadrupole mode.

The primary component of EL CMi is the third object in which pulsation modes that occur only in triaxially distorted stars have been reported. In that light it is logical to suspect that the previously discovered tidally tilted or single sided pulsators may be just special cases of triaxial pulsators. After all, all of these objects are distorted into triaxial shapes, and their pulsation modes conform to the same framework. The single mode of HD 74423 is consistent with a tidally enhanced $Y_{10x}$ mode, whereas the four modes of CO Cam appear to be of the same type. The behavior of the single mode of TIC 63328020, originally classified as a tidally tilted $l=1$, $m=1$ ($Y_{11x}$) mode, is also consistent with a $Y_{10y}$ mode. The numerous mode doublets and one triplet in TZ Dra \citep{2022MNRAS.510.1413K} are also consistent with $Y_{10x}$ and $Y_{10y}$ modes, respectively, and the pulsation modes of the subdwarf B star HD 265435 \citep{2022ApJ...928L..14J} conform to the predictions of the behavior of modes in triaxially distorted stars by \cite{2025ApJ...979...80F} as well.

As regards to the evolutionary state of the tidally tilted or triaxial pulsators with a main sequence star as the primary components, EL CMi, TIC 63328020 and TZ Dra have undergone or are undergoing mass transfer, whereas HD 74423, CO Cam, TIC 184743498 and TIC 435850195 are detached systems. Thus it appears that the appearance of these special types of pulsation is not connected to a particular evolutionary phase of close binary systems.

The pulsating star in EL CMi is a very simple and clear example that demonstrates that the main features of the theoretical description of ``Fuller modes" are consistent with observations. It is reasonable to expect that in the future, stars with richer, more complicated pulsation spectra will be discovered. Given that these pulsations in essence come with an observational identification of the underlying modes, and occur in stars that can be characterized with a high degree of confidence and accuracy given that they are located in binary and multiple systems, these objects could be used for detailed asteroseismic studies, using oscillations that take place around three different symmetry axes. In other words, such objects would hold the promise of studying the interior structure of distant stars in three dimensions.

\begin{acknowledgements}
\noindent 
GH and AM both thank the Polish National Center for Science (NCN) for supporting this study through grant 2021/43/B/ST9/02972 and Filiz Kahraman Ali\c{c}avu\c{s} for her comments on the spectrum of the target.

DJ acknowledges support from the Agencia Estatal de Investigaci\'on del Ministerio de Ciencia, Innovaci\'on y Universidades (MCIU/AEI) under grant ``Nebulosas planetarias como clave para comprender la evoluci\'on de estrellas binarias'' and the European Regional Development Fund (ERDF) with reference PID-2022-136653NA-I00 (DOI:10.13039/501100011033). DJ also acknowledges support from the Agencia Estatal de Investigaci\'on del Ministerio de Ciencia, Innovaci\'on y Universidades (MCIU/AEI) under grant ``Revolucionando el conocimiento de la evoluci\'on de estrellas poco masivas'' and the the European Union NextGenerationEU/PRTR with reference CNS2023-143910 (DOI:10.13039/501100011033). H-LC is supported by the National Key R\&D Program of China (grant Nos. 2021YFA1600403), the National Natural Science Foundation of China (grant Nos. 12288102, 12333008 and 12422305). VK acknowledges support from NSF grant AST-2206814.

This paper includes data collected by the TESS mission. Funding for the 
TESS mission is provided by the NASA Science Mission Directorate.
The QLP data used in this work were obtained from MAST
(\url{https://dx.doi.org/10.17909/t9-r086-e880}),
hosted by the Space Telescope Science Institute (STScI).
STScI is operated by the Association of Universities for 
Research in Astronomy, Inc., under NASA contract NAS 5–26555.

This work also presents results from the European Space Agency (ESA) 
space mission Gaia. Gaia data are being processed by the Gaia Data 
Processing and Analysis Consortium (DPAC). Funding for the DPAC is 
provided by national institutions, in particular the institutions 
participating in the Gaia MultiLateral Agreement (MLA). The Gaia 
mission website is \url{https://www.cosmos.esa.int/gaia}. The Gaia archive
website is \url{https://archives.esac.esa.int/gaia}. This research has 
also made use of the VizieR catalogue access tool, CDS, 
Strasbourg, France. 
\end{acknowledgements}

\section*{Data Availability}
We make all files needed to recreate our \textsc{mesa-binary} results publicly available at Zenodo: \url{https://zenodo.org/records/15745684}

\section*{Software}
This paper made use of the following codes/packages:
         \textsc{lightkurve} \citep{lightkurve},
         \textsc{IRAF} \citep{iraf}
\textsc{SPECTRUM} \citep{1994AJ....107..742G}
\textsc{Period04} \citep{2005CoAst.146...53L}
\textsc{MESA} \citep{Paxton2011, Paxton2013, Paxton2015, Paxton2018, Paxton2019, Jermyn2023}

\bibliography{elcmi}{}
\bibliographystyle{aasjournal}

\end{document}